# Characterisation of slip and twinning in high rate deformed zirconium with electron backscatter diffraction

**Vivian Tong*[1], Euan Wielewski[2], T. Ben Britton[1]**

1. Department of Materials, Imperial College London, Exhibition Road, London, SW7 2AZ, UK

2. School of Engineering, University of Glasgow, University Avenue, Glasgow, G12 8QQ, UK

*Corresponding author: v.tong13@imperial.ac.uk

## Abstract

Zirconium alloys are used in the nuclear industry as structural materials, and can be subject to high strain rate loading conditions during forming and in the case of a reactor accident. In this context, the relationship between strain rate dependent mechanical properties, crystallographic texture and deformation modes, such as slip and deformation twinning, are explored in this work. Commercially pure zirconium is deformed to 10 % engineering strain under quasi-static and high strain rate loading, and post-mortem analysis of the samples is performed using electron backscatter diffraction (EBSD) to observe different twin and slip systems activated. Twin types are identified from local intergranular misorientation maps, and active slip systems are identified from long range intragranular misorientation maps. We link characterisation of the mechanical responses, twin types and morphologies, and relative slip system activation as a function of loading mode. We find that variations in strength and hardening can be related to the relative propensity of twinning and the number of active slip systems.

# Graphical Abstract

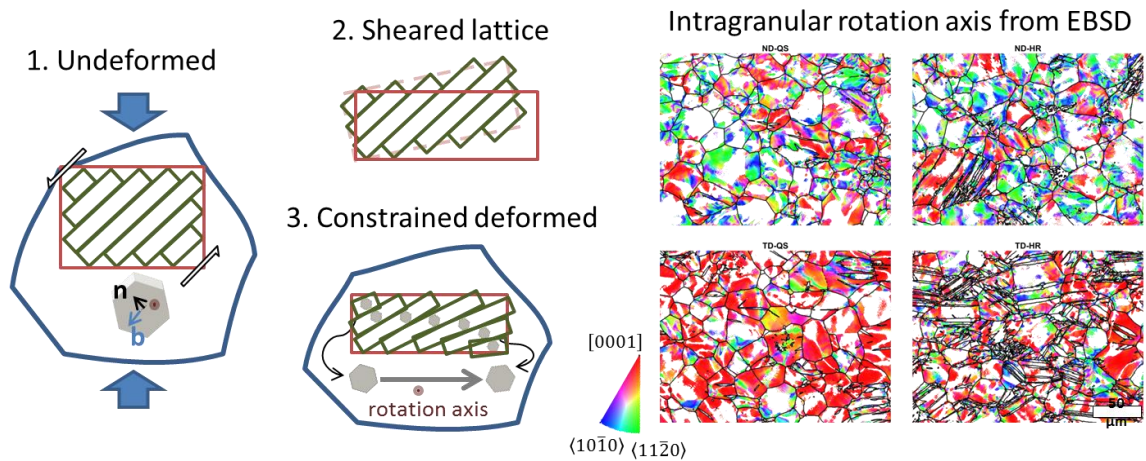

# 1 Introduction

Zirconium alloys are used in nuclear reactors as fuel rod cladding due to zirconium's high strength and low neutron absorption cross section. Knowledge about the rate dependence of competing deformation modes is important, especially during forming of the cladding tubes and during reactor accident conditions. Zirconium has a hexagonal close packed crystal structure at room temperature, where $\langle a \rangle$ prismatic slip has the lowest critical resolved shear stress. $\langle a \rangle$ slip is orthogonal to the unit cell $\langle c \rangle$ axis and therefore cannot accommodate deformation along $\langle c \rangle$ [1]. To make up the five independent slip modes and allow arbitrary deformation in a polycrystal, secondary deformation systems such as twinning along pyramidal planes and $\langle c + a \rangle$ slip on either $1^{st}$ order or $2^{nd}$ order pyramidal planes play an important role in Zr polycrystal deformation. Therefore, the relative activity of deformation slip and twinning modes as a function of texture and strain rate is critical in understanding deformation behaviour. Anisotropic deformation during processing affects texture of the final Zr part, understanding the relative predominance of deformation twinning and slip is important for both texture control in processing and predicting likely failure modes in-service.

Electron backscatter diffraction (EBSD) is a technique used to map lattice orientation (and phase) across a crystalline sample. Grain maps, and identification of twins in these maps, can be readily achieved and residual twin content can be assessed.

Use of EBSD, along with cross-correlation based, high angular resolution EBSD, has been used to explore the consequences of plastic strain, and for instance it has been used to estimate geometrically necessary dislocation (GND) densities [2–7] and total dislocation densities [8] in polycrystalline materials. This method fits the local lattice rotation gradients in small kernels from 2D crystal orientation maps to the local rotation fields expected from a network of stored dislocations. Measurement of residual dislocation density can be used to understand the deformation behaviour, but they are only a signature of the deformation history, and it is not always possible to link GND density to plastic strain, as (for instance) single slip deformation in a relatively unconstrained area does not result in local changes in lattice curvature [9]. For controlled deformation histories, and with calibration, it can be possible to find correlations between dislocation density and plastic strain level during multiple slip deformation in polycrystals, as dislocation cell structures develop with increasing plastic strain [10,11]. In contrast, twinning is easier to observe with EBSD, as the twin misorientation angle/axis and twin shape are directly measured [12–14]. In EBSD maps of deformed metals, there are often long-range orientation gradients that extend across a grain. In this work, we propose evaluation of the axis and angle of these longer range orientation fields to explore the spatial activity of crystal slip, when a grain is contained within a constrained polycrystalline network.

## 1.1 Deformation slip

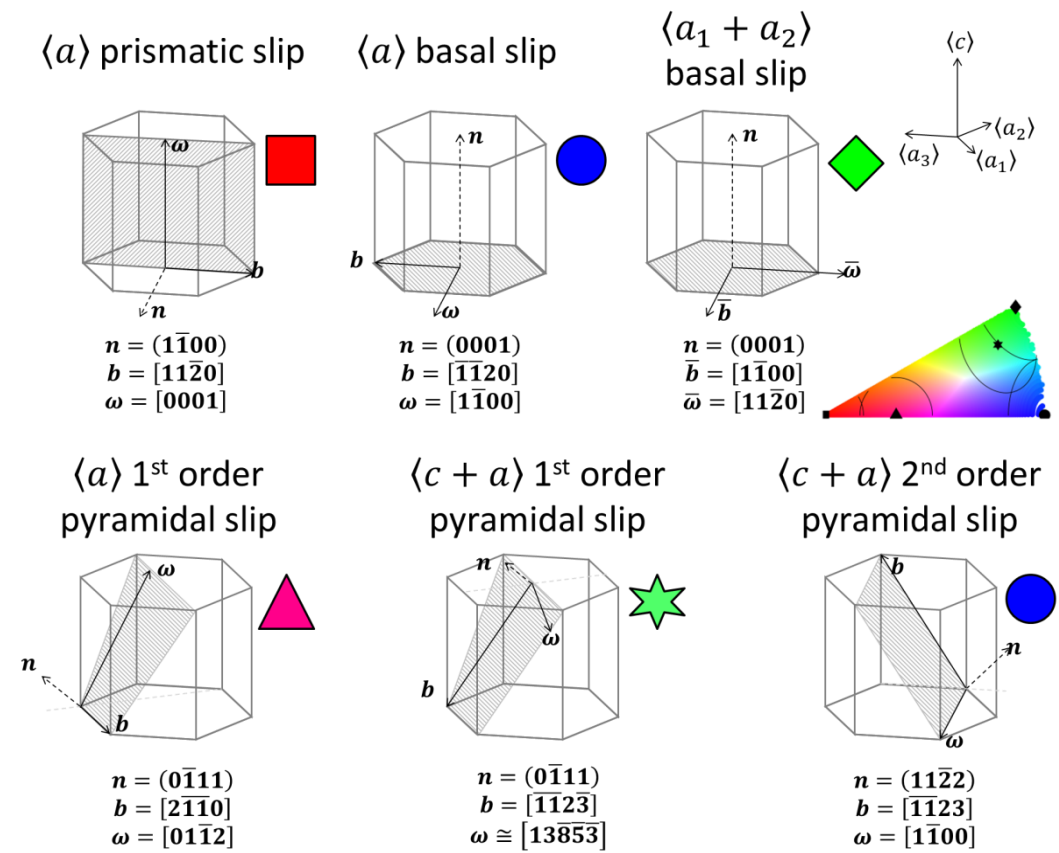

**Figure 1:** Slip systems in zirconium alloys. $\boldsymbol{b}$ and $\boldsymbol{n}$ are the slip direction and slip plane respectively [20,41,52,53], and $\boldsymbol{\omega}$ is the rotation axis calculated in the present work, orthogonal to both the slip plane normal and slip direction. The crystal direction of the rotation axis vectors are labelled on the IPF colour key.

The known deformation systems in Zr are shown in Figure 1. The preferred room temperature slip system with lowest critical resolved shear stress (CRSS) in dilute Zr alloys is $\langle a \rangle$ prismatic slip [15]. The CRSS of $\langle a \rangle$ prismatic slip increases with interstitial content, notably oxygen, carbon and nitrogen, and decreases with increasing temperature [16]. $\langle a \rangle$ basal slip in high purity single crystal Zr deformed at a low strain rate of $10^{-4}$ s$^{-1}$ was only seen at temperatures above 550°C [17]. At room temperature basal slip is seen to occur in small amounts as a secondary slip system to $\langle a \rangle$ prismatic slip, and is promoted during high strain rate loading [18,19]. In room temperature deformation studies of Zr, $\langle a \rangle$ basal slip is sometimes ignored [20,21] and has been shown not to affect macroscopic stress-strain response at room temperature [22]. However, single crystal room temperature microcantilever tests in commercial purity Zr show that $\langle a \rangle$ basal slip has only 1.3 times higher CRSS than $\langle a \rangle$ prismatic slip, which would imply significant activation in polycrystal deformation given a favourable stress state. $1^{st}$ order $\langle c + a \rangle$ pyramidal slip has a 3.5 times higher CRSS than $\langle a \rangle$ prismatic slip [15]. Slip on $2^{nd}$ order pyramidal planes is rarely seen in Zr alloys, but $\langle c + a \rangle$ $1^{st}$ order pyramidal slip is commonly observed [15,23–25]. Jensen and Backofen [1] observed localised shear bands with $\langle c + a \rangle$ dislocations on $\{11\bar{2}4\}$ planes during $\langle c \rangle$ axis loading, which led to ductile fracture at room temperature, but this is not the slip plane as $\langle c + a \rangle$ vectors do not lie in $\{11\bar{2}4\}$ planes.

Deformation twinning produces a coordinated shear transformation in a crystalline material. Twin types can be classed as either contraction (C1, C2) or extension (T1, T2) twins, which accommodate strain either to contract or extend the <c> axis of the hexagonal close-packed (HCP) unit cell.

Twinning is crystallographically defined by its twin plane $\boldsymbol{K_1}$, the mirror plane in the twin and parent material, and $\boldsymbol{\eta_1}$ which is the twinning shear direction. Deformation twins in Zr are generally lenticular in shape, lengthening in the $\boldsymbol{\eta_1}$ direction and thickening along the $\boldsymbol{K_1}$ plane normal [26].

The twin plane, shear direction, and shear plane form the basis vectors of an orthogonal set. The axis-angle misorientation relationship between the parent and twin is a rotation of angle $\xi$ about the shear plane normal direction $\boldsymbol{P}$.

More generally, twinning can be described as a 180° rotation about an axis ($\eta_1$ or $K_1$ normal direction), or a mirror reflection in a plane ($K_1$ or $\eta_1$ normal plane). The predominant twin type in zirconium is $K_1 = \{10\bar{1}2\}\ \eta_1 = <10\bar{1}1>$ (T1) twinning and for this $\{10\bar{1}2\}<10\bar{1}1>$ twin there is no distinction between the four transformations, as they are equivalent [26].

The four major twin types in zirconium are listed in Table 2.

Due to symmetry in the HCP crystal structure, there are six crystallographically equivalent twin variants for each twin type. Different twin variants of the same type in a grain cannot be distinguished by their axis-angle disorientation to the parent, which are the same for all variants of a twin type, but they can be distinguished apart using their absolute orientations with respect to the loading axis, and in some cases (depending on the sectioning plane) the twin boundary trace.

The primary twin type formed in any sample depends on the strain state and rate, temperature and crystal orientation. In macroscopic samples this is typically influenced strongly by the crystallographic texture, grain size, competing deformation modes (i.e. dislocation slip), combined with the loading axis and direction. The T1 twin type is dominant at room temperature and quasi-static strain rates [21]. Twin types present at liquid nitrogen temperature are $\{11\bar{2}2\}\langle11\bar{2}\bar{3}\rangle$ (C1 twinning) and $\{10\bar{1}2\}\langle10\bar{1}1\rangle$ (T1 twinning). Secondary twins of another type may form inside the primary twins as the crystal is reoriented with respect to the loading axis [21,27]. The C2 compressive twin system $\{10\bar{1}1\}\langle\bar{1}012\rangle$ is only active at high temperatures [28,29], and is activated in preference to basal slip during deformation at 550°C [17].

## 1.2 Influence of loading conditions on deformation modes

Kaschner and Gray [30] observe that yield stress increases with increasing strain rate in the range of 0.001 s$^{-1}$ and 3500 s$^{-1}$, and that the strain rate sensitivity in the yield stress is higher when uniaxially compressing along texture components with predominantly prismatic planes than basal planes. They conclude that the rate sensitivity of the flow stress is consistent with Peierls forces inhibiting dislocation motion in low-symmetry metals during slip-dominated deformation. This is valid in the early stages of room temperature deformation, which in Zr is usually slip-dominated [31].

Samples compressed along texture components with predominantly prismatic planes yield at lower stresses than texture components with predominantly basal planes [30], consistent with the higher critical resolved shear stress for $<c + a>$ pyramidal slip compared to $<a>$ prismatic slip. In a TEM study of room temperature deformed zirconium, McCabe et al. [21] observed only $<a>$ dislocations in samples with prismatic texture, which were presumed to lie on prismatic planes. Both $<a>$ (prismatic) and $<11\bar{2}\bar{3}>$ $<c + a>$ ($\{10\bar{1}1\}$ pyramidal) slip were observed in samples with basal texture at room temperature, but only $<a>$ dislocations were observed in the same sample at liquid nitrogen temperature.

At quasi-static strain rates, McCabe et al. [21] only observed T1 twinning in samples compressed along a plate direction with a prismatic texture component along the loading axis. They did not observe T1 twinning in samples compressed along basal textures to 25% strain. Kaschner and Gray observe that deformation at high strain rates (3000s$^{-1}$) produces more twins than at quasi-static strain rates, but the twin types activated were not identified [30].

Capolungo et al. [14] studied twinning as a function of grain orientation within a sample. They calculated a global Schmid factor using the macroscopic applied stress direction, and found the resolved shear stress on any grain without taking into account local intergranular interactions which may alter the actual stress state. They found that although the majority of twins occur in grains favourably oriented for twinning according to the global Schmid factor, around 30% of grains which were unfavourably oriented for twinning still contained twins. Likewise, the twins present were not always of the highest global Schmid factor variant, with only 60% of twins twinning on the highest Schmid factor variant. This can be attributed to a strong dependence on the local stress conditions in grains or grain boundaries [32], which is difficult to measure experimentally, particularly at high strain rates.

Knezevic et al [22] fitted experimental data of high purity polycrystalline Zr to a viscoplastic self-consistent model to study the rate and temperature sensitivity of slip and twinning systems. They found that T1 twinning was the dominant slip system at room temperature for strain rates between $10^{-3}$ and $10^{3}$ s$^{-1}$. Basal slip did not contribute to deformation below 400°C. Twinning was found to be rate insensitive and changes in twinning behaviour as a function of strain rate could be explained by the rate sensitivity of slip.

## 1.3 Identification of slip activity

The main methods used to identify slip system activity involves either slip trace analysis of single crystals [15,33] or polycrystals [34], diffraction techniques such as neutron diffraction [19] and high angular resolution electron backscatter diffraction elastic strain analysis [35], or TEM diffraction imaging of dislocations [21]. These methods are summarised in Table 1 with respect to discrimination of slip systems in zirconium.

In slip trace analysis, only the slip plane is measured, and the slip direction is inferred. In zirconium, this enables identification of slip activity on a basal, prism, or 1st/2nd order pyramidal plane. In the case of a 1st order pyramidal plane trace, slip could be in either $\langle a \rangle$ or $\langle c + a \rangle$ directions; slip trace analysis cannot discriminate between these.

Diffraction-based studies measure the residual dislocation content as opposed to the slipped dislocations, which is only a good approximation for systems which accumulate networks of geometrically necessary dislocations, such as face centre cubic polycrystals [11]. In low-symmetry crystals such as hexagonal zirconium, there could be regions of predominantly single slip where geometrically necessary dislocations may not necessarily accumulate [9]. Residual dislocation content does not distinguish between glissile and sessile dislocations. Glissile dislocations contribute to slip and hardening, but sessile dislocations contribute only to latent hardening.

The slip plane of a residual dislocation cannot generally be resolved by diffraction methods. In Zr, the screw components of $\langle a \rangle$ dislocations could slip on any of the prismatic, basal, or 1st order pyramidal planes. Similarly, $\langle c + a \rangle$ screw dislocations could slip on either 1st or 2nd order pyramidal planes.

Our work is on characterisation of deformation using EBSD, which include slip and twinning in Zr. In particular, we have developed a method to identify slip system activity from long range intragranular lattice rotation axes, when in a constrained slip geometry. Our method is complementary to conventional slip analysis methods as it can identify slip systems which tend to be missed using other methods.

# 2   Method and materials

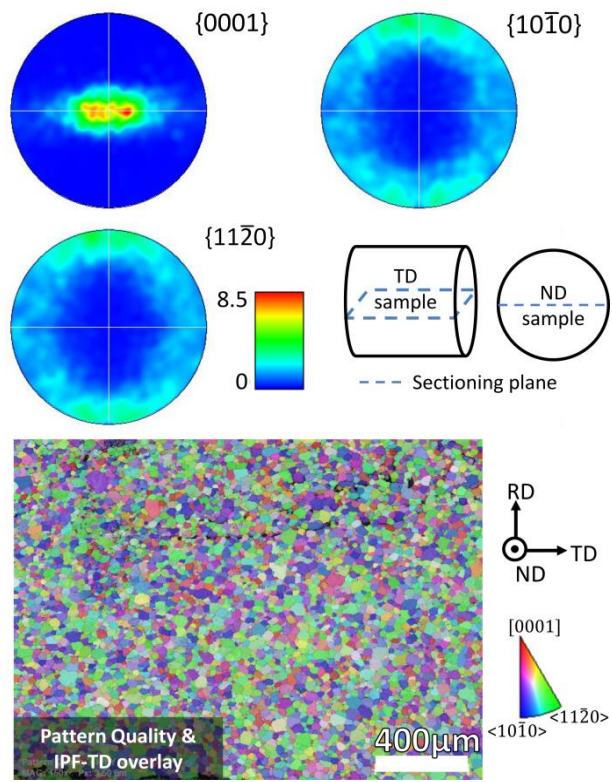

**Figure 2:** Texture and grain morphology of starting material, with pole figure colour scale in multiples of random orientation. The average grain size measured by EBSD is 25µm.

The starting material used in this study was commercially pure zirconium (CP-Zr) of 97.0% purity, from Advent Research Materials Ltd. The texture measured by electron backscatter diffraction (EBSD) in Figure 2 shows that the as received material had a typical unidirectionally rolled and recrystallised texture, indicating that the final rolling pass dominated the texture evolution during processing. Therefore, plate directions will be referred to as the rolling/normal/transverse directions (RD/ND/TD). Pole figures and an EBSD IPF colour map shows the texture of the material measured from ~6000 grains with average grain size of 25 µm (Figure 2).

Four cylindrical mechanical testing samples (diameter 3.0 mm and height 3.5 mm) were machined from the as-received plate, two with the cylinder axis aligned along the TD direction ('TD samples') and two along ND directions ('ND samples'). Samples along each direction were compressed to 10 % engineering strain in either a split-Hopkinson bar [36] at a high strain rate of $10^3\,s^{-1}$ ('HR samples'), with a copper ring based momentum trap, or in a conventional crosshead device for mechanical testing at a quasi-static strain rate of $10^{-3}\,s^{-1}$ ('QS samples'). A detailed overview of the experimental procedure used for the quasi-static and high strain rate mechanical testing is given by Wielewski et al. [37]. The anisotropic deformation behaviour in this same material has also been characterised by Taylor impact testing by Wielewski et al. [38].

The deformed samples were cross-sectioned, ground to 4000 grit (5 μm particle size) with SiC papers, and then electropolished for 120 seconds at 25 V and -40 °C in an electrolyte solution of 50ml perchloric acid and 450 ml methanol. EBSD data of the deformed samples were collected on either a Zeiss Auriga field emission gun SEM or a FEI Quanta field emission gun SEM. In both cases, data was collected using a Bruker eFlashHR EBSD camera with a step size of 0.3 μm over a 240×180 μm area.

The fields of view were chosen to be approximately in the centres of the sample to minimise observation of edge friction artefacts during compression. Noting from prior work by Morrow et al. [39] that use of a different sectioning plane does not significantly skew twin statistics, and that microstructural variability can present a bigger problem, larger EBSD maps (which are not shown here) were also collected for all samples to verify that there were no significant variations in texture or twinning frequency along the compression axes and sample radii.

The orientation data was post-processed using in-house software written in MATLAB to produce grain boundary maps with twin types identified. Details of the post-processing algorithms can be found in Reference [40]. In all maps the indexing rate exceeded 95 %, and two iterations of data clean up were used for the grain boundary maps so that twin boundaries could be properly identified. The custom data clean up algorithm dilates the EBSD map into unindexed points, assigning these points the average neighbour orientation.

Twin boundaries were identified using the quaternion representation of the four common twin misorientations [28], which can be characterised by a rotation about the shear plane normal axis. The intragranular misorientations for the different twin types in zirconium are tabulated in Table 2 as axis-angle pairs. From the twin and grain boundary edge maps, the twin numbers, lengths and widths were measured by hand. Area fractions were calculated from segmentation of grain and twin boundary maps with twins manually highlighted by flood filling. As twinning can fragment the original grains, nearby grains of similar orientation (mean orientations < 4° apart) were grouped and treated as a single grain during postprocessing.

Likely slip systems activated were analysed by mapping intragranular misorientation axes and angles, calculated from the misorientations between orientation quaternions of each data point and its grain mean. Misorientation axis directions are plotted in both sample reference frame (orthorhombic symmetry) and crystal reference frame (hexagonal symmetry).

As misorientation angle decreases, degeneracy in the rotation axis leads to a large measurement uncertainty. Therefore, crystal rotation axes for the slip systems plotted in the inverse pole figure in Figure 1 have 15° loci around the ideal rotation axes to reflect this. Further justification for this is given in the Appendix.

# 3 Results

## 3.1 Mechanical data

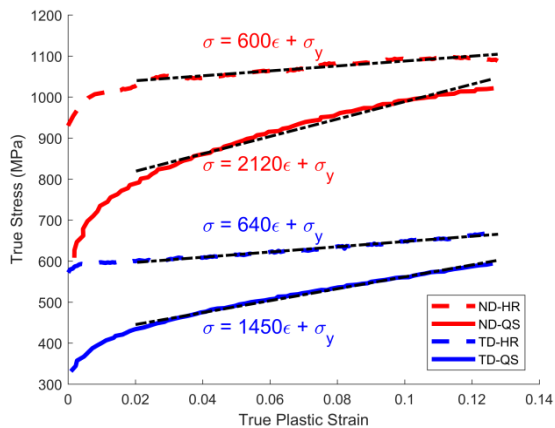

**Figure 3:** Mechanical data of commercially pure Zr in compression. ND (red) – normal direction, TD (blue) – transverse direction, HR (dashed lines) – high strain rate ($10^3 s^{-1}$), QS (solid lines) – quasi-static strain rate ($10^{-3} s^{-1}$). Black dashed lines are linear best fits.

Figure 3 shows the mechanical response of the samples during deformation, as true stress plotted against true strain. TD loaded samples are softer than ND loaded samples, and quasi-static strain rate deformed (QS) samples are softer than high rate deformed (HR) samples. Approximate work hardening rates were calculated by least squares fitting of straight lines to the data (black dashed lines).

The HR samples harden less strongly than the QS samples. Between the QS samples, the TD loaded sample hardened less than the ND loaded sample. The hardening rates of the two HR samples are similar. The flow stresses are strongly dependent on texture and, to a lesser extent, increase with increasing strain rate.

## 3.2 Texture and Twins

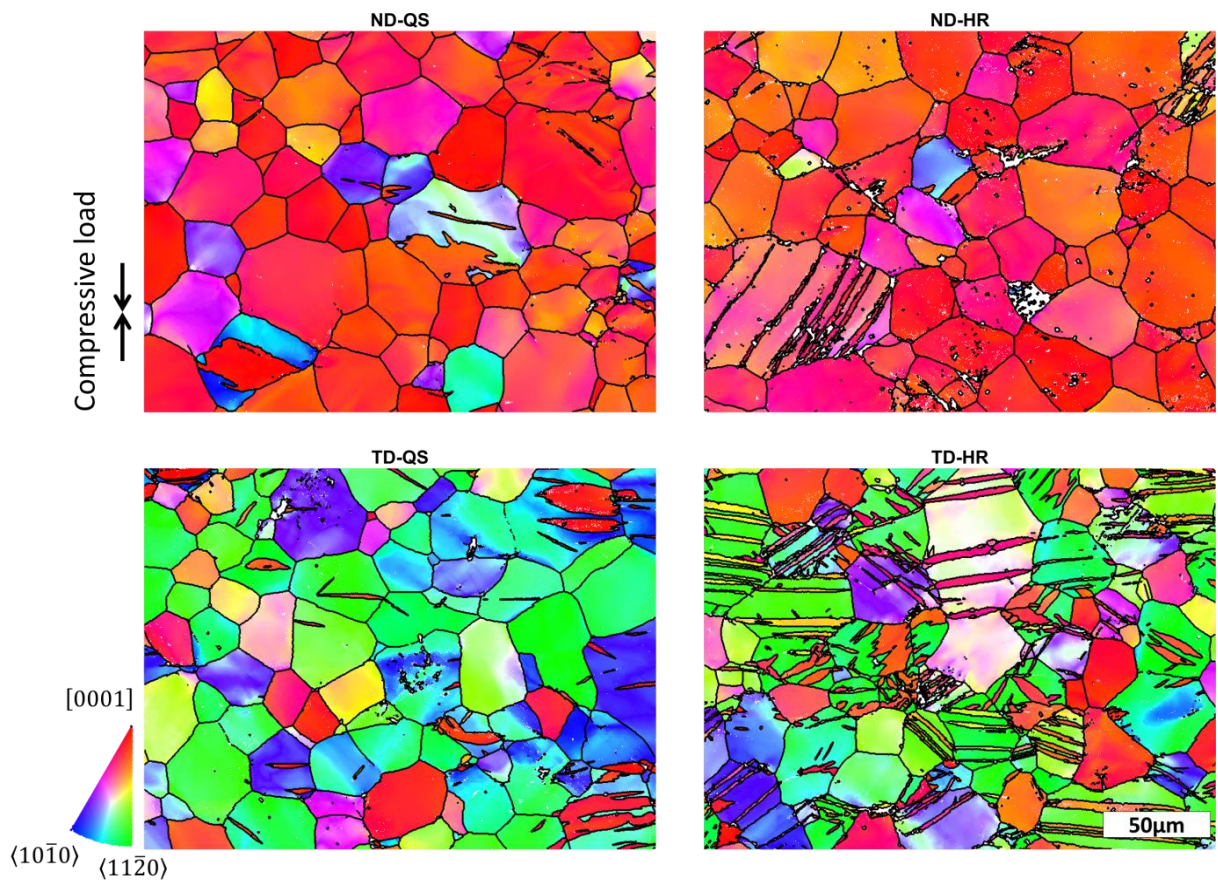

**Figure 4:** EBSD IPF maps along the macroscopic compressive loading direction overlaid with grain boundary type. The loading direction is vertical in the maps. The horizontal axis points along RD in all samples. All boundaries are outlined in black.

Figure 4 shows the crystal orientations of points within the deformed samples coloured according to inverse pole figure (IPF) maps of the four samples with respect to the loading axis. All grain boundaries are coloured in black and unindexed points are white.

There is a texture difference between the ND and TD samples, which matches the pole figure in Figure 2. In the ND samples, the $\langle c \rangle$ directions (red) are aligned approximately along the loading axis. In the TD samples, the prismatic directions of most grains are aligned along the loading axes, apart from in twins, where the crystal lattice has been rotated away from the parent orientation. Sample TD-HR in particular has twinned extensively.

## 3.3 Intergranular and Intragranular Misorientation Angles

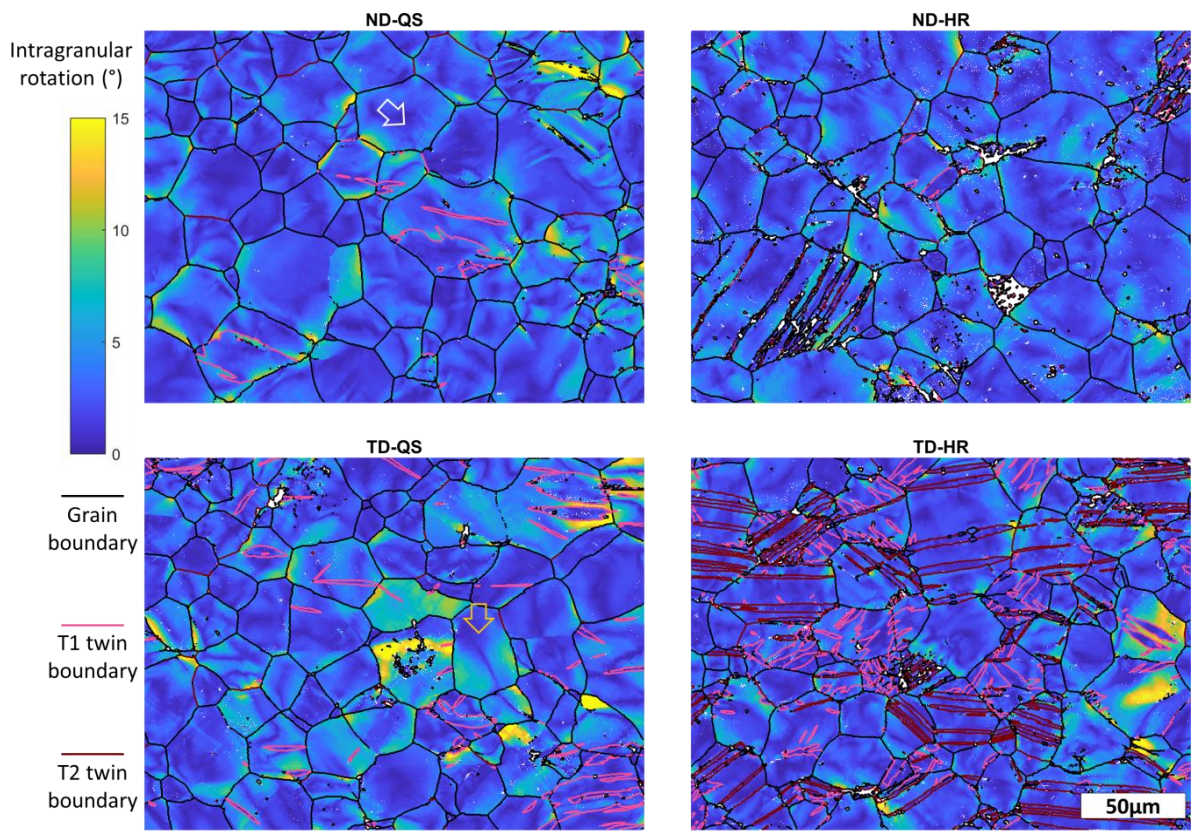

**Figure 5:** Intragranular misorientation (GROD) angles overlaid with grain boundary type. Non-special grain boundaries are outlined in black, T1 twin boundaries in pink, and T2 twin boundaries in brown.

Figure 5 shows EBSD maps of intragranular misorientation angle at each point relative to the grain mean orientation, also known as grain reference orientation deviation (GROD). In the limit that the grains are constrained by their neighbourhood, these maps can be interpreted to estimate shear strain in a constrained polycrystal.

The intragranular rotation fields are heterogeneous in all samples, and systematic differences between samples are not obvious. TD-HR has slightly lower average misorientation in the grains, most likely because the majority of deformation is carried by twinning instead of slip. The misorientations are often larger near grain boundaries and triple junctions. Misorientation angles within twins are smaller than both parent grains and untwinned grains in all samples.

The rotation fields are either smoothly increasing to large rotation angles near grain boundaries (e.g. yellow arrow in TD-QS) or smaller magnitude 'flecks' at a length-scale comparable to the EBSD step size of 0.3µm (e.g. white arrow in ND-QS). Predominantly single slip is expected in regions with smoothly varying rotation fields. Multiple slip activation leading to dislocation cell structure formation is expected in regions containing smaller flecks, which are of similar lengthscale to dislocation cell structures seen in Zr (~0.2 µm cell size at 10-20 % strain [21,31]).

Local intergranular rotations are shown at grain boundaries and coloured according to boundary type, as twin boundaries can be identified by specific misorientation angles and axes, given in Table 2. Non-special grain boundaries are highlighted in black, T1 twin boundaries in pink, and T2 twin boundaries in brown. A few isolated grain boundary segments coloured in brown (e.g. the top right region of ND-QS) are not twin boundaries but have a similar misorientation to T2 twin boundaries. (As significant plastic slip had occurred after twin growth in some grains in ND-HR, the maximum misorientation deviation from ideal twin relation had to be relaxed to 12° to identify the twins sufficiently, as shown in Supplementary Figures 1-2.)

T1 twins (pink boundaries) are present all samples to varying extents, but T2 twinning (brown boundaries) is only activated in high rate deformation. In all samples, most of the twins have $\langle c \rangle$ axis aligned near the loading axis, which is a hard (high flow stress) orientation (Table 3). This means that minimal slip deformation should be expected within twins, though parent grain fragments may continue to slip after twinning.

Table 3 shows manually measured grain and twin statistics for the four samples. Additional mapping performed on as received material showed that the starting average grain diameter was 25μm (Figure 2). The grain size broadly decreases with increasing twinning, as the twins fragment grains. At this strain level, no grain size or grain reorientation hardening effect, typified by a concave stress-strain curve [21,22,31], can be seen in the mechanical data (Figure 3).

T1 twins have a lower aspect ratio than T2 twins, which tend to elongate along the twinning shear direction until blocked on both ends by grain boundaries, instead of thickening normal to the twinning plane. In contrast, many T1 twins are pinched and terminate at grain interiors, and sometimes grow into non-lenticular shapes which consume much of the parent grains. T2 twins on average are larger than T1 twins. At this strain level, a similar area fraction of T1 and T2 twins are nucleated in the HR samples.

The plastic strain from twinning can be calculated from $\varepsilon_{twin} = \sqrt{\frac{1}{2}} s V_t$, where $s$ is the magnitude of twinning shear for each twin type (0.167 for T1, 0.63 for T2), and $V_t$ is the volume fraction of twins [28,31]. In this data set, we estimate the plastic strain from twinning can be calculated by approximating the twin area fraction to be similar to the volume fraction.

In the ND-QS, ND-HR and TD-QS samples, twinning shear leads to 1% plastic strain or less, accommodating around 8% of the total plastic deformation (12.5% true strain). This implies that in these samples, most of the deformation has been accommodated by slip. In the TD-HR sample, which has twinned significantly more, twinning shear leads to 6% plastic strain, i.e. twinning carries around half the total deformation (13%, as per Figure 3).

## 3.4 Intragranular misorientation axes

### 3.4.1 Misorientation axes plotted in crystal reference frame

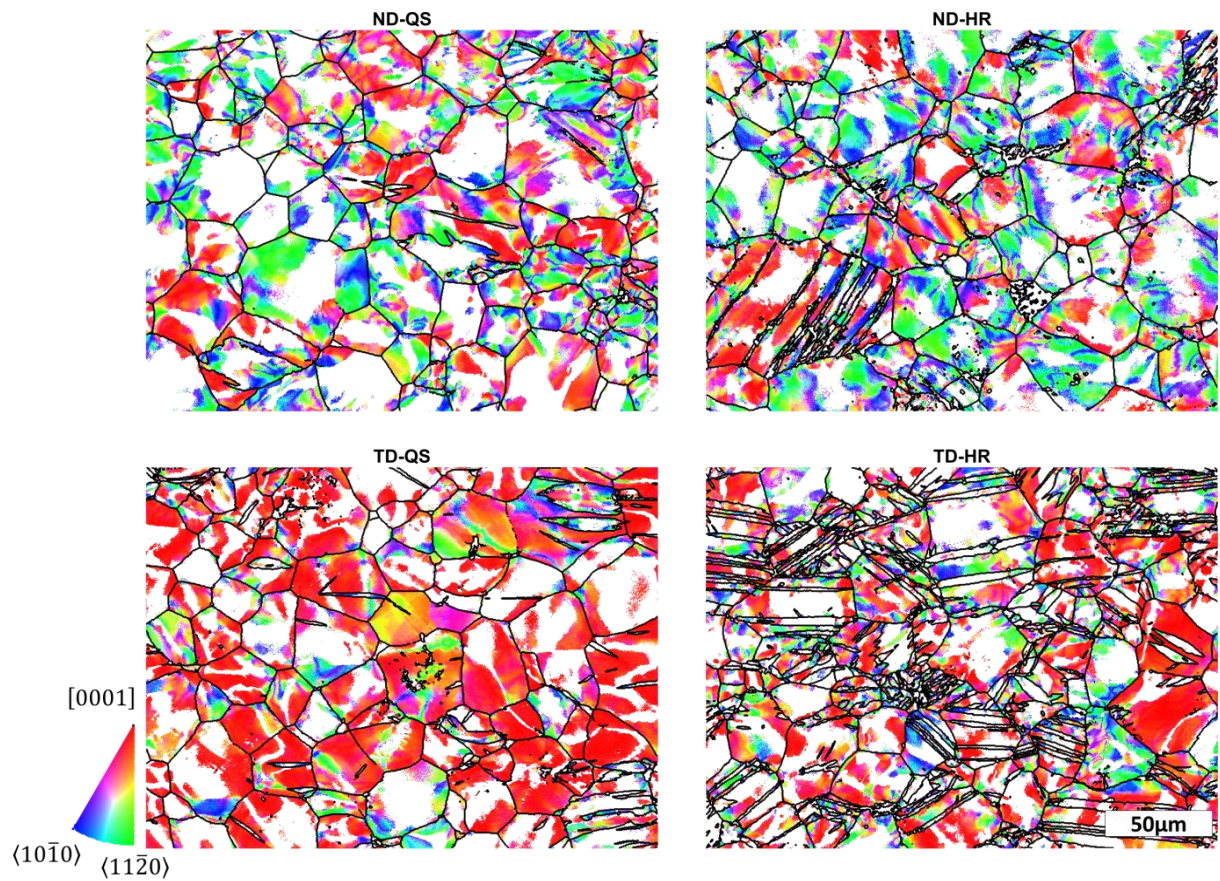

**Figure 6:** Intragranular misorientation (GROD) axes in crystal reference frame overlaid with grain boundary type. Misorientation axis vectors are plotted in hexagonal symmetry IPF colouring, showing the rotation axis as a direction in the HCP unit cell. All boundaries are outlined in black.

The misorientation axes in Figure 6 are plotted with respect to the average orientation of the grains, i.e. crystal directions in the final grain configuration.

The crystal directions of the misorientation axes show the crystal rotation axis during slip deformation. In single slip, the rotation axis is normal to both the slip plane normal and the slip direction of the active dislocations. The expected rotation axis, and its colouring, are shown in Figure 1. A $[0001]$ rotation axis corresponds to $\langle a \rangle$ prismatic slip, which is red in the IPF map. A $\langle 10\bar{1}0 \rangle$ rotation axis corresponds to $\langle a \rangle$ basal slip, which is blue in the IPF map. A $\langle 11\bar{2}0 \rangle$ rotation axis (green in IPF map) corresponds to a special case of multiple slip, where two adjacent ⟨a⟩ basal slip systems are activated in similar proportions in the same region [41]. The white regions in Figure 6 are areas with misorientation angles of less than 2°, which have a large uncertainty in the misorientation axis measurement (see Appendix).

In QS deformation, the TD sample shows predominantly red and the ND sample a mixture of blue, green and red colours. During HR deformation, blue/green colours increase and red colours are reduced. The prevalence of each slip system is quantified in Table 4.

### 3.4.2 Misorientation axes plotted in sample reference frame

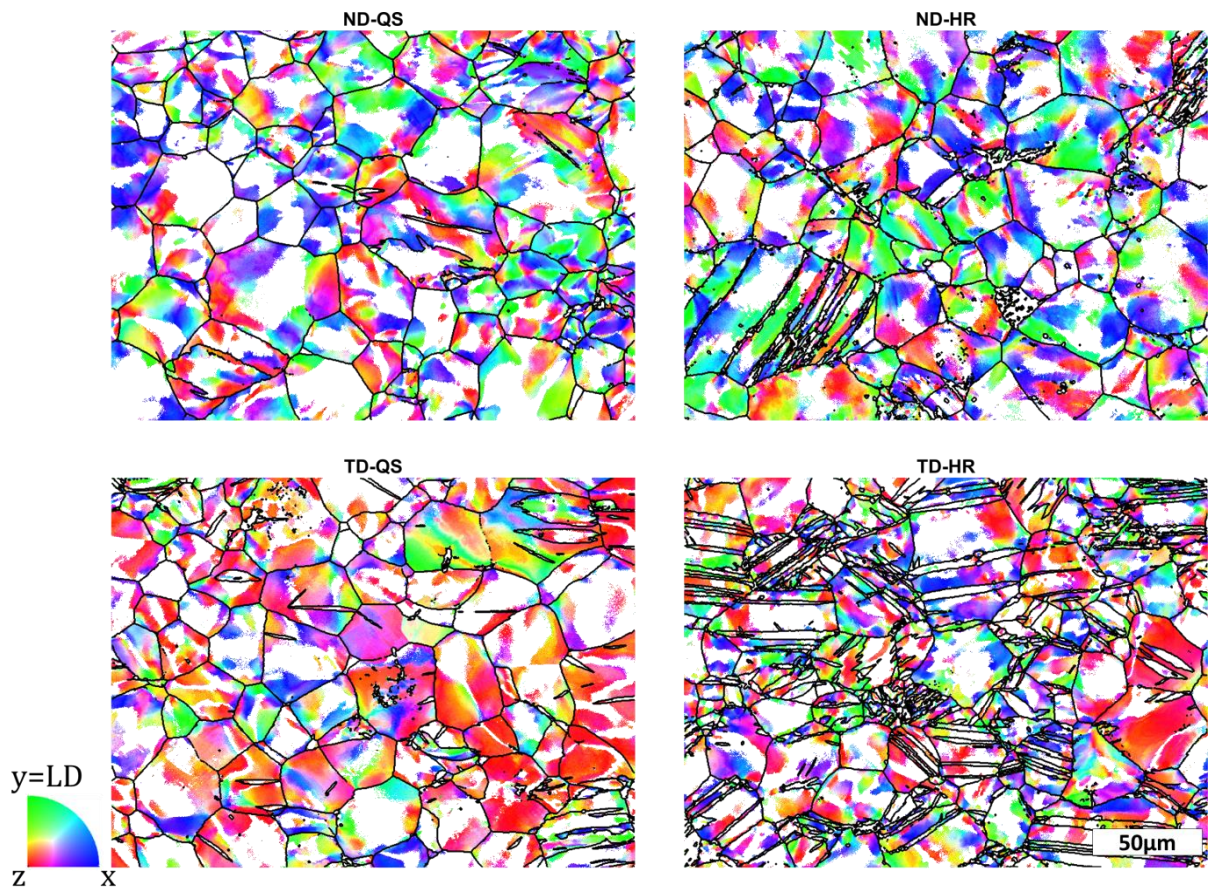

**Figure 7:** Intragranular misorientation (GROD) axes in sample reference frame overlaid with grain boundary type. Misorientation axis vectors are plotted in orthorhombic symmetry IPF colouring, showing the rotation axis as a direction in the sample. All boundaries are outlined in black.

The misorientation axes in Figure 7 show lattice rotation axes plotted with respect to the sample axes. The IPF colours are qualitatively different from sample to sample because of the different textures and slip systems activated. For all slip systems apart from ⟨a⟩ prismatic slip, different symmetric variants of the slip system will show up as different colours. Variants of $\langle a \rangle$ prismatic slip are not distinguishable in the sample frame GROD axis, as the rotation axis of $\langle a \rangle$ prismatic slip is [0001] for all slip variants. The misorientation axis maps in sample frame identify regions of uniform slip by a common colour. This is subject to bias in colour perception, as visually obvious colour transitions in the IPF key do not necessarily correspond to large misorientations, and vice versa.

# 4 Deformation mechanisms

## 4.1 Lattice rotation axes for hexagonal close-packed slip systems

The $\langle a \rangle$ basal and $\langle c + a \rangle$ 2$^{nd}$ order pyramidal slip systems have identical rotation axes, and cannot be distinguished by GROD axis. $\langle a_1 + a_2 \rangle$ double basal slip is not distinguishable from $\langle c + a \rangle$ 1$^{st}$ order pyramidal slip, though there are points which can be unambiguously assigned to $\langle c + a \rangle$ 1$^{st}$ order pyramidal slip. Since $\langle c + a \rangle$ 2$^{nd}$ order pyramidal slip is rarely seen in zirconium alloys, it is likely that the blue regions in Figure 6 are mostly $\langle a \rangle$ basal slip, not $\langle c + a \rangle$ 2$^{nd}$ order pyramidal slip.

## 4.2 ⟨a⟩ prismatic slip

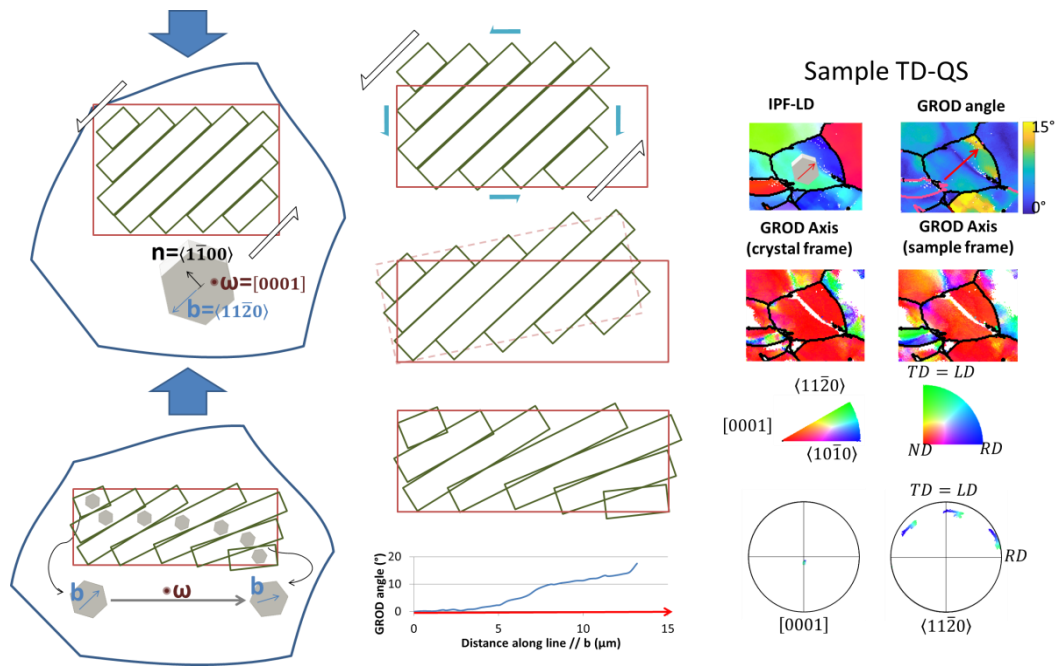

**Figure 8:** Schematic showing grain rotation from single slip in a constrained geometry. An example grain from sample TD-QS deformed by single ⟨a⟩ prismatic slip is shown.

Figure 8 schematically shows how lattice rotation occurs following single activation in a constrained geometry. Shear in the slip direction $b$ along the slip plane $n$ leads to a plastic rigid body rotation which must be balanced by (elastic) lattice rotations to maintain the constrained grain geometry.

An experimentally measured case of ⟨a⟩ prismatic slip in sample TD-QS is demonstrated.

The misorientation axis is [0001] (red in crystal frame), corresponding to ⟨a⟩ prismatic slip activation, and a uniform colour in the sample frame, which shows that no other slip system is not activated (such as ⟨a⟩ pyramidal, which has a crystal rotation axis close to ⟨a⟩ prism, as seen in Figure 9). The grain orientation pole figures show the [0001] grain rotation axis in single slip, which has preserved its initial orientation, and a rotation gradient in the slip direction ⟨11$\bar{2}$0⟩.

The slip is presumed to be single slip, although this cannot be determined from the sample frame GROD axis in the case of $\langle a \rangle$ prismatic slip. This assumption is probably correct, as the Schmid factor for the most favourable variant of $\langle a \rangle$ prismatic slip is 0.47, and the other two variants are considerably lower at 0.29 and 0.11. The GROD angle monotonically increases along the presumed favourable slip direction, which matches the expected behaviour for single slip shown in the schematic.

Even in single slip, the degree of slip is modulated by its neighbours – in Figure 8, regions of very high misorientation angle are seen at the north-east boundary between the central grain and its hard neighbour, which is not deforming by $\langle a \rangle$ prismatic slip. Slip has localised in the north-east boundary region because it is deformed in preference to its hard neighbour poorly oriented for slip. The other neighbours have relatively soft orientations and mostly deformed by $\langle a \rangle$ prismatic slip, allowing more uniform slip and a smaller rotation gradient near the grain boundary.

## 4.3 Distinguishing slip variants in adjacent regions

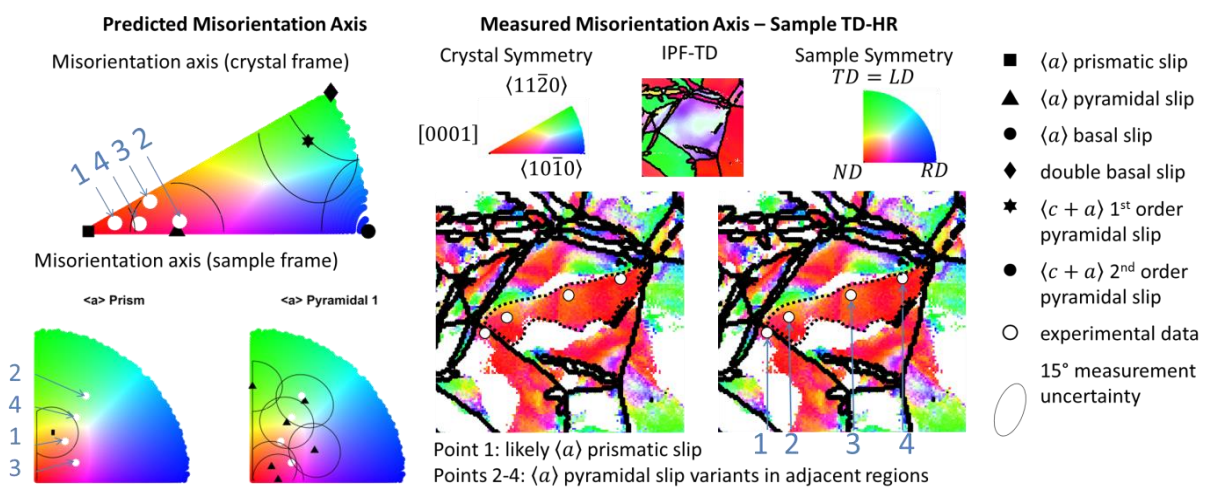

**Figure 9:** An area from sample TD-QS showing adjacent regions of uniform slip, with different variants of $\langle a \rangle$ pryamidal slip activated.

Figure 9 shows an area in sample TD-HR with adjacent regions of $\langle a \rangle$ prismatic and $\langle a \rangle$ pyramidal slip. The misorientation axes in crystal symmetry show that the same slip system family is activated, and the misorientation axis in sample symmetry shows patches of similar colour as regions of uniform slip.

Four points in this region of the microstructure, labelled points 1-4, are plotted as white circles in Figure 9. The expected misorientation axes in the crystal frame for each slip system (as shown in Figure 1) are compared with the experimentally measured data. Within the 15° uncertainty of the axis measurement, points 2, 3 and 4 correspond to $\langle a \rangle$ pyramidal slip and point 1 corresponds to $\langle a \rangle$ prismatic slip.

The rotation axes for $\langle a \rangle$ pyramidal and $\langle a \rangle$ prism slip are also plotted in the sample frame for this grain mean orientation and compared to experimentally measured misorientation axes. Point 1 matches misorientation axes for both $\langle a \rangle$ prismatic and $\langle a \rangle$ pyramidal slip in the sample frame IPF, but matches only $\langle a \rangle$ prismatic slip in the crystal frame IPF. Therefore, it is likely to have undergone $\langle a \rangle$ prismatic slip. Points 2-4 correspond to three different variants of $\langle a \rangle$ pyramidal slip respectively, although which slip variant point 3 belongs to is ambiguous due to the overlap in uncertainty bounds of the slip variants.

In this grain, it is possible that $\langle a \rangle$ screw dislocations slipped on $\langle a \rangle$ prismatic and $\langle a \rangle$ pyramidal planes in adjacent regions.

## 4.4 Basal slip and $\langle c + a \rangle$ pyramidal slip systems

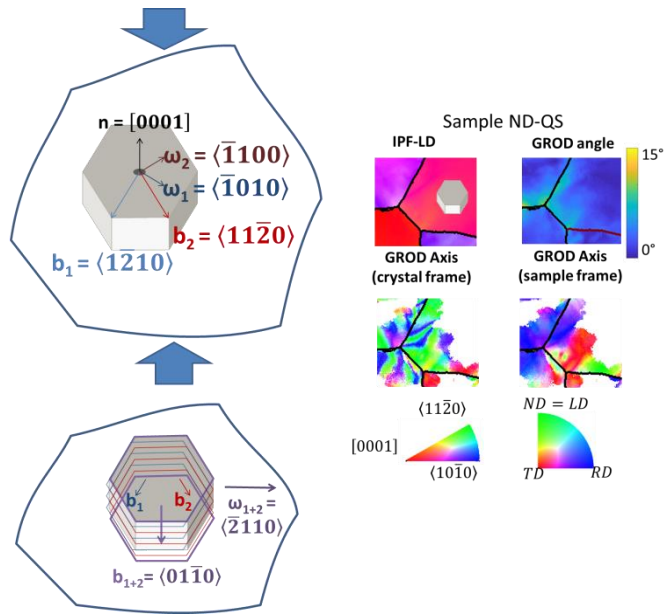

**Figure 10:** Schematic showing grain rotation from two $\langle a \rangle$ basal slip variants leading to double basal slip, a special case of multiple slip. An example grain from sample ND-QS is shown alternating between $\langle a \rangle$ basal and double basal slip.

The misorientation axis IPF maps in crystal frame (Figure 6) often have regions of green and blue next to each other. This is especially prevalent in the ND-HR map. Since the green and blue regions often occur next to each other, the dislocation types are probably related, paired either as $\langle c + a \rangle$ 1$^{st}$ order pyramidal adjacent to 2$^{nd}$ order pyramidal dislocations, or (perhaps more likely, due to a lower CRSS) two $\langle a \rangle$ basal slip systems adjacent to $\langle a_1 + a_2 \rangle$ basal dislocations.

Figure 1 shows that blue regions with $\langle 10\bar{1}0 \rangle$ type rotation axes correspond to $\langle a \rangle$ basal slip or $\langle c + a \rangle$ 2$^{nd}$ order pyramidal slip activation. It is not possible to distinguish these two slip systems as they share a rotation axis. Green regions with $\langle 11\bar{2}0 \rangle$ type rotation axes correspond to either $\langle a_1 + a_2 \rangle$ double basal slip or $\langle c + a \rangle$ 1$^{st}$ order pyramidal slip. Although the rotation axes for these two slip systems are distinct, the IPF key in Figure 1 shows that $\langle a_1 + a_2 \rangle$ double basal slip is completely contained within the uncertainty limits of $\langle c + a \rangle$ 1$^{st}$ order pyramidal slip.

$\langle a_1 + a_2 \rangle$ double basal slip is a cooperative slip mechanism where two variants of $\langle a \rangle$ dislocations are activated on the same basal plane, shifting the rotation axis away from $\langle 10\bar{1}0 \rangle$. In the limit, two adjacent $\langle a \rangle$ basal dislocation types contribute equally to slip deformation, leading to a pseudo-single slip mode (double basal slip) with a $\langle 11\bar{2}0 \rangle$ type rotation axis (green in the IPF map). Figure 10 shows how equal activation of two adjacent $\langle a \rangle$ basal slip systems can lead to double basal slip. In polycrystal deformation, the relative proportions of two active basal slip systems can change, so the rotation axis alternates between $\langle 11\bar{2}0 \rangle$ (double basal slip, IPF green) and $\langle 10\bar{1}0 \rangle$ (single basal slip, IPF blue).

## 4.5 Multiple slip

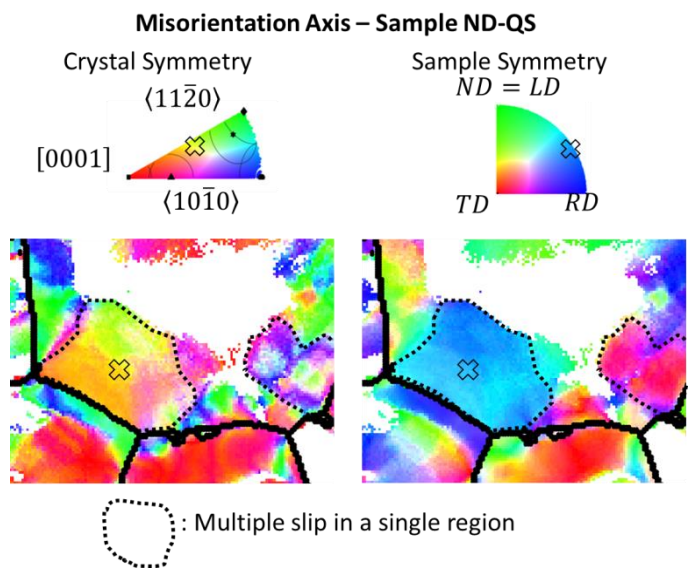

**Figure 11:** An area of sample ND-QS showing a relatively uniform region of multiple slip.

Two regions of multiple slip are highlighted by dotted lines in Figure 11. The left region, a region of cooperative multiple slip, has a cross in the middle which is marked out in the IPF keys by corresponding crosses. The right region shows multiple impinging slip systems activated on a longer lengthscale.

The left region shows a region of uniform slip, with uniform misorientation axis IPF colour in both sample and crystal frame. However, the misorientation axis does not correspond to any Zr slip system in the crystal frame IPF key, therefore there must be multiple slip systems active in this region which contribute to the grain rotation axis. Since the misorientation axis is relatively uniform, the slip systems must be acting cooperatively to rotate the crystal in a single direction, on a lengthscale smaller than the 0.3 µm EBSD step sized used here.

The right region shows many rotation axes in the crystal frame IPF map, but a relatively uniform colour in the sample frame IPF map. This region shows many slip systems activated on a lengthscale that is resolved in this EBSD map. The dislocation cell structure in this region is also visible in the GROD angle map (shown in Figure 5, white arrow in sample ND-QS) as flecks of varying GROD angle.

Generally, in multiple slip activation, the total long range rotation axis is the combination of rotation caused by many slip systems which may switch on and off at different stages, and locations, during deformation. For an ideal set of orientation data (with no measurement uncertainty) it would be possible to tell whether multiple slip occurred at a given point, but it is not possible to deconvolute this into which types of dislocations were activated during the multiple slip deformation, and at what stage they were activated, without prior knowledge of the initial configuration (this is not trivial experimentally).

## 4.6 Work hardening

A few mechanisms contribute to work hardening active at this strain level: During single slip activation, dislocations can pile up at obstacles, such as grain boundaries (self-hardening); or interact with inactive dislocations (forest hardening, or latent hardening). Single slip (Stage II work hardening) leads to a constant work hardening rate and linear stress-strain response. Multiple slip activation, either homogeneously in the entire grain or from local intersection of single slip regions, causes dislocation cell structure formation. Multiple slip (Stage III work hardening) leads to a decreasing work hardening rate and parabolic stress-strain response [42] . Twinning causes grain refinement and lattice rotation to harder orientations with higher flow stresses. The effect of lattice rotation is particularly pronounced in HCP Zr due to its significant slip system anisotropy [28].

Experimental work hardening behaviour can tell us, on average, which deformation mechanisms are active in the samples. The mechanical data shows good linear fits (black dashed best fit lines in Figure 3) for all samples but ND-QS, where the experimental data significantly deviates from the linear fit. This means that multiple slip is expected in sample ND-QS whereas majority single slip is expected in the other three samples.

12 % true strain is relatively early in the deformation process, so we do not expect all active hardening mechanisms to be shown in the mechanical data. For example, there is considerable grain fragmentation from twinning in TD-HR (as seen later in Figure 4), but the stress-strain curve is not sigmoidal and therefore this data does not show the typical response of a grain size hardening effect [21,22,31]. The sigmoidal stress-strain curve arising from further slip in twin-fragmented grains only appears after around 30 % true strain in Zr-1Nb [43] and 15 % true strain in pure Zr [31], depending on when during the deformation process twinning deformation is exhausted and the twinned grains start to slip. Therefore, at 12.5 % true strain, a pronounced sigmoidal shape due to twinning in TD-HR is not expected, nor is it seen in the mechanical data in Figure 3.

# 5 Discussion

## 5.1 Analysis method

### 5.1.1 Twinning

Slip first, then twinning results in a twin with a near-ideal twin misorientation. Twinning first, followed by slip results in a twin with a non-ideal twin misorientation. The large grain with long and wavy T2 twins in sample ND-HR is misoriented far from the ideal T2 twin, which is why only part of the boundary points are identified as T2 twin boundary points, even when the maximum misorientation deviation (usually the Brandon criterion for cubic metals [44]) from the ideal twin relation is relaxed to 12° (see Supplementary Figures 1-2).

### 5.1.2 Grouping of parent grain fragments

GROD measures misorientations with respect to the grain average orientation, which in the present work is used as a proxy for the initial grain orientations. As twinning fragments the parent grains, it should be considered whether the mean orientation should be calculated separately for each grain fragment, or if the fragments should be grouped when calculating mean grain orientation. We have performed the calculations for both cases, and presented GROD maps with respect to the mean parent orientation of all fragments.

Grouping of grain fragments is required if the majority of slip occurred before twinning. If the grain slipped as a single unit before twinning, deformation in the parent fragments should be measured with respect to a single reference. If significant slip occurred before grain fragmentation from twinning, the GROD will be underestimated if the parent fragments are not grouped.

Similarly, if twinning fragmented the grains early in the deformation, so that the grain fragments slip as separate units, the GROD could be overestimated if the parent fragments are grouped. Grouping of parent grain fragments for calculation of grain mean orientation makes a difference to the absolute values in the GROD angle maps. It does not affect qualitative trends or distributions.

The stage of deformation at which twinning occurred can be estimated by the deviation of twin boundary misorientations from the ideal twin relations given in Table 2. If significant slip happened after twinning, the twin misorientation will be far from ideal, and vice versa. The potential GROD error from grouping or not grouping grains is < 5° in the current datasets (see Supplementary Figure 3).

The GROD axis is negligibly affected, as the measurement uncertainty of 15° is large compared to the difference grouping makes to grain mean orientations (< 5°).

### 5.1.3 Slip system determination from GROD misorientation axis

#### 5.1.3.1 Method

Shearing of a crystal lattice through slip leads to plastic rotation of the object. This plastic rotation must be balanced by the lattice rotation gradient if the system is constrained to maintain grain compatibility. Subject to this assumption of constraint, the lattice rotation gradient is related to the amount of plastic shear.

As an estimate, it is possible to calculate a balance of plastic and elastic (lattice) rotations of a constrained object in simple shear shows that the typical 10° of rotation over a length (grain radius) of 15 μm corresponds to 8000 $\langle a \rangle$ slip steps at the grain boundary, as shear strain is defined as the tangent of the plastic rotation angle in a simple shear geometry, which in a constrained geometry is balanced by the elastic lattice rotations (due to compatibility).

$$\gamma_{plastic} = -\gamma_{elastic} \qquad \text{Equation 1}$$

$$\tan 10° = |\gamma_{plastic}| = |\gamma_{elastic}| = \frac{slipped\ length}{grain\ radius} \qquad \text{Equation 2}$$

$$\tan 10° = \frac{slipped\ length}{15000\ nm} \qquad \text{Equation 3}$$

$$slipped\ length \cong 2600 nm \cong 8000\ \langle a \rangle \qquad \text{Equation 4}$$

This is different to a dislocation density measurement, as dislocation density counts the residual dislocation content present after deformation. Most of these dislocations would not be visible during e.g. transmission electron microscopy, but for some crystal orientations can be measured by the height of slip steps after deformation. The calculation here also assumes a homogeneous shear and lattice rotations with minimal localisation, which is not always a good assumption near grain boundaries, as seen in Figure 5.

Lattice rotation in the crystal is relative to the initial undeformed grain configuration. In our analysis, we use the grain mean orientation to approximate this when plotting the angle, as the initial orientations were not measured, which is not necessarily correct (see Figure 8 for a schematic of grain rotation during single slip where the average orientation is not preserved). Depending on the shape of the rotation field, the true rotation angles (with respect to the initial configuration) may be more than double that of the reported rotation angles (with respect to the grain mean). However, at this macroscopic strain level (12.5 % true strain), the average grain rotation due to slip is likely to be relatively small, as evidenced by the deformed textures being similar to the undeformed pole figures in Figure 2 apart from extra orientations present from twinning.

*5.1.3.2 Application*

Where two variants of the same slip system family are activated in adjacent regions, they are identified as different colours in Figure 7, but would be masked by crystal symmetry operators in the IPF maps of Figure 6 where the axes are plotted as crystal directions. Figure 11 shows a case of a single region of multiple slip activation, where the misorientation axes have uniform colour in sample frame but do not correspond to a single dislocation type in crystal frame. In contrast, Figure 9 shows adjacent regions of $\langle a \rangle$ prismatic and pyramidal slip. It is not obvious in the crystal frame that there are multiple $\langle a \rangle$ pyramidal slip variants activated as the entire region is red, but regions of different colours when plotting in sample frame show this area has multiple variants activated. The slip variant analysis does not extend to $\langle a \rangle$ prismatic slip, which has a [0001] rotation axis. As there is only one [0001] direction in the HCP unit cell, the three $\langle a \rangle$ prismatic slip variants are not distinguishable by their misorientation axis in the sample frame.

## 5.2 Comparison of hardening data with deformation mechanisms

Slip contributes to work hardening by dislocation-dislocation interactions (multiple slip only) or dislocation-obstacles, such as-grain boundary interactions (single slip and multiple slip), whereas twinning does not harden by dislocation interactions but can reorientate the material into a harder orientation, and thus reduce the mean free path for mobile dislocations and therefore refine the grain size. Coherent twins can also nucleate dislocations at the twin boundary due to the lattice mismatch strain, and the extent of this hardened region is a function of the twin thickness [45].

Table 3 shows that in samples ND-QS, ND-HR, and TD-QS, only ~8% of the total plastic deformation is accommodated by twinning. Sample TD-HR accommodates about half of the applied deformation by twinning. The concave stress strain response characteristic of twin-dominated deformation is expected only at higher (typically ~30%) strains [31].

Single slip (hardening by local compatibility constraint and latent geometric hardening) causes linear work hardening, whereas multiple slip (compatibility constraint, latent hardening, and slip system interactions) leads to parabolic work hardening behaviour. In quasi-static loading, higher prevalence of multiple slip correlates with non-linear hardening: ND-QS, with a measured multiple slip fraction of 22% (Table 4), shows non-linear work hardening; TD-QS shows less multiple slip (15%, Table 4) and relatively linear hardening. This trend does not apply to the high rate deformed samples, but the comparison is more complicated as ND-HR is slip-dominated but TD-HR has significant contribution from twinning.

## 5.3 EBSD based observation of basal and 1st order pyramidal slip

Existing literature suggests that basal slip is anomalous at these strain levels in Zr literature [18] but reported to be replaced by T1 or C2 twinning and $\langle a \rangle$ prismatic slip at 550°C and below [17], and only significant at room temperature for 50% strain and above when work hardening exhausts further slip on $\langle a \rangle$ prismatic slip systems [46]. In these prior studies, the interstitial solute content, which can dramatically affect CRSS, is typically very low (120 ppm or under of oxygen), differing from the low purity starting material (1600 ppm O) used in this work. Our results show that for this material and texture (typical of unidirectionally rolled Zr) it dominates in room temperature ND compression both at QS and HR strain rates. This matches micro-cantilever testing observations of material with similar oxygen content, which places the CRSS of $\langle a \rangle$ basal slip as 1.3 times that of $\langle a \rangle$ prismatic slip [15].

$\langle a \rangle$ 1$^{st}$ order pyramidal slip is found to be an important slip mode in all samples, activated more than $\langle a \rangle$ prismatic slip in TD samples, but it is rarely reported in Zr literature. The prevalence of $\langle a \rangle$ pyramidal slip could be interpreted as $\langle a \rangle$ screw dislocations slipping on 1$^{st}$ order pyramidal planes. In conventional analysis methods $\langle a \rangle$ pyramidal dislocations would not be distinguishable from other $\langle a \rangle$ screw dislocations in TEM analysis of residual dislocations, or 1$^{st}$ order pyramidal slip planes (with $\langle c + a \rangle$ slip direction assumed) in slip trace analysis. The prevalence of $\langle a \rangle$ 1$^{st}$ order pyramidal slip matches observations in HCP Ti, where Naka et al. found that single crystals oriented favourably for $\langle a \rangle$ prismatic slip cross slip onto 1$^{st}$ order pyramidal planes due to a non-planar dislocation core structure which dissociates into two partial dislocations which can cross-slip easily onto the 1$^{st}$ order pyramidal plane [47].

## 5.4  Effect of loading mode on slip

Table 4 shows qualitative estimations of the spatial frequency of slip for the four samples, allowing for 15° deviation from the ideal misorientation axis.

This analysis does not account for all slip activity, as we do not consider points with a small misorientation (<2°) and there is some overlap between $\langle a \rangle$ prismatic and $\langle a \rangle$ pyramidal slip. Overlapped points were counted twice, but the overlapped fraction is low, as seen in the bottom two rows of Table 4. Furthermore, Table 4 shows the spatial frequency of slip and but does not show the degree of slip in a region, which is related to GROD angle.

GROD axis vectors corresponding to $\langle a_1 + a_2 \rangle$ basal slip are completely contained within the uncertainty bounds of $\langle c + a \rangle$ 1$^{st}$ order pyramidal slip, but there are points which can be unambiguously identified as $\langle c + a \rangle$ 1$^{st}$ order pyramidal slip. Therefore, points which could belong to either slip system has been reported as $\langle a_1 + a_2 \rangle$ basal slip only, which is likely as the CRSS for basal slip is lower than for $\langle c + a \rangle$ $1^{st}$ order pyramidal slip at room temperature quasi-static loading [15].

For the TD samples, $\langle a \rangle$ prismatic and $\langle a \rangle$ pyramidal slip dominate, and in the ND samples $\langle a \rangle$ basal slip. At high rate $\langle a \rangle$ prismatic slip is suppressed for both texture components, and $\langle a \rangle$ basal slip is promoted. This matches quasi static micro-cantilever measurement of similar Zr material, where the CRSS for $\langle a \rangle$ basal slip is 1.3 times that of $\langle a \rangle$ prismatic slip, which is accounted for by differences in flow stress (Figure 3). Misorientation axes not corresponding to any single slip system ('Multiple Slip' in Table 4) is also a significant fraction. Though multiple slipped regions can be identified on a case by case basis such as in Figure 11, this is not trivial in an automated sense and so in Table 4 they are joined together.

The small GROD angles in twins compared to in parent grains (Figure 5) suggests that there is minimal slip inside twins. This is not surprising, as the twinning observed here generally reorients the crystal lattice to a harder slip orientation. However, this may be an artefact of the much smaller grain width of the twins: Intragranular misorientation angles estimate the slipped length as they lead to a rotation in a constrained geometry; therefore equivalent strain in a smaller grain will lead to a shorter slipped length and consequently a smaller rotation angle measured.

## 5.5 Effect of loading mode on twinning

T1 twinning is activated in all samples: both ND and TD textures at both QS and HR strain rates. T1 twinning is more prevalent in TD samples where the average grain $\langle c \rangle$ axis is near perpendicular to the loading direction. In ND samples it is only nucleated in grains with $\langle c \rangle$ axis oriented away from the loading axis. T2 twinning is activated only in HR samples, and carries much more of the plastic deformation than T1 twinning in these samples (Table 3).

There is a large difference in contribution to strain in both HR samples, despite similar area fractions of T1 and T2 twins, because T2 twins have a much higher twinning shear than T1 twins. The high T2 twinning shear causes an large elastic backstress on the parent grain [48,49], which explains the tendency for T2 twins to preferentially lengthen instead of thicken, and for many thin twins of few variants to form in the same grain. By the same argument, as T1 twins have a low twinning shear and little backstress in the parent grain, they can thicken easily, many nucleated variants can grow, and are even reported to consume entire grains [50]. The relationship between high twinning shear and thin twins matches observations and interpretations in very early work by Rapperport [51], although they were confused because their T2 twinning shear was incorrect (their reported T2 twinning shear is 0.216 [51], as opposed to the correct value of 0.63 [52]).

# 6  Conclusions

This paper describes a method to measure the activity of twinning and slip systems during polycrystal deformation from post-mortem EBSD maps and applies this to polycrystalline commercially pure Zr.

T1 twinning occurs during both quasi-static and high rate loading. T2 twinning occurs only at high rate loading. Similar area fractions of T1 and T2 twinning are activated at high strain rate, but T2 twinning carries more of the plastic deformation due to its higher twinning shear. T1 twins tend to thicken with incoherent boundary traces in preference to lengthening along the twinning plane, and in some cases nearly consume the entire parent grain. Several variants of T1 twins can nucleate in the same grain and the twin tips are pinched at grain interiors. On the other hand, T2 twins preferentially lengthen instead of thicken, and tend to nucleate in parallel rows of the same variant extending from boundary to boundary.

GROD axis analysis was used to identify slip systems activated by assuming a geometrically constrained polycrystalline system, and using the grain mean orientation as a proxy for undeformed orientations. For this Zr composition, basal, $\langle a \rangle$ pyramidal and $\langle c + a \rangle$ pyramidal slip systems dominate room temperature compression along ND at both quasi-static and high strain rate loading, which is not seen in high purity polycrystalline and single crystal Zr. In $\langle a \rangle$ axis (TD) deformation, $\langle a \rangle$ prismatic and $\langle a \rangle$ pyramidal slip systems are dominant. $\langle a \rangle$ pyramidal and basal slip systems are more prevalent than currently reported in the literature, though this may be because $\langle a \rangle$ pyramidal slip is not easily identified by conventional analysis routes.

Basal slip systems are promoted and $\langle a \rangle$ prismatic slip is suppressed at high strain rate (HR) compared to quasi-static strain rate (QS) loading. This is independent of loading axis texture (ND/TD).

# 7 Acknowledgements

The authors would like to thank Clive Siviour for providing material and access to mechanical testing facilities. TBB and VT would like to thank EPSRC (EP/ K034332/1) for funding on the HexMat Programme Grant (http://www.imperial.ac.uk/hexmat). TBB would like to thank the Royal Academy of Engineering for funding his Research Fellowship.

EW performed the mechanical testing. VT performed the microscopy. TBB wrote the initial software used for EBSD data postprocessing from which VT developed the GROD axis analysis. TBB supervised the work.

# 8  Appendix

Uncertainty in the misorientation axes is limited by the angular resolution in EBSD measurement, as measured by Wilkinson [54]: although the misorientation angle is well defined for rotation angles as small as 0.33°, scatter in the misorientation axis is considerably larger: for misorientations angles of 2°, the mean error in the rotation axis is around 15°. With intragranular rotations we necessarily deal with small angles with potentially large errors.

In Figure 6, misorientation axes of points with GROD values above 2° are plotted. Below this misorientation angle, the error in the misorientation axes is too large to unambiguously identify slip systems.

Another uncertainty to consider in the GROD misorientation axis is the deviation between grain mean orientation, which the misorientations are calculated with respect to, and the grain initial orientation, which gives the true rotation axis but is not measured. In the case of a single slip system active in the entire grain, a large deviation from the initial grain orientation does not necessarily lead to uncertainty in the misorientation axis. The direction of the rotation axis is not changed or smeared out by the grain rotation, as shown in the [0001] pole figure in Figure 8, so the misorientation axis is correct with respect to both the initial and final orientation.

However, in grains with multiple regions of uniform slip (e.g. Figure 9), error in the grain mean orientation will lead to errors in the misorientation axis with magnitudes on the order of the GROD angles. The maximum likely error arising from a poor approximation of the grain initial orientation is around twice the 95$^{th}$ percentile GROD angle, which is 7.4° for these data. This is around the same as the EBSD misorientation axis measurement error (around 15°).